\documentclass[letterpaper, 10 pt, conference]{ieeeconf}  
\IEEEoverridecommandlockouts                              
\usepackage{epstopdf} 
\usepackage{amsmath} 
\usepackage{amssymb}  
\title{\LARGE \bf
	{An I + PI Controller Structure for Integrating Processes with Dead-Time: Application to Depth Control of an Autonomous Underwater Vehicle}
}
\author{Sujay D. Kadam$^{1}$\\ SysIDEA Lab, IIT Gandhinagar
	\thanks{$^{1}$Sujay D. Kadam ({\tt\small kadam.sujay@iitgn.ac.in}) is a doctoral student in Electrical Engineering at SysIDEA Lab, IIT Gandhinagar, India.        }%
}
\usepackage{epstopdf,xcolor,setspace,epsfig}
\usepackage[utf8]{inputenc}
\usepackage[T1]{fontenc}
\let\labelindent\relax
\usepackage{longtable,breakcites,enumitem} 
\usepackage{amssymb,amsmath,amsbsy,amsfonts,graphicx}
\usepackage{mathptmx}
\usepackage{balance}
\usepackage[cal=boondoxo]{mathalfa}
\usepackage{hyperref}
\newcounter{defcounter}
\usepackage{ifpdf}
\begin{document}
\maketitle
\begin{abstract}
The paper presents a feedforward plus feedback controller structure with I and PI controllers for control of an integrating process with dead time. Guidelines for controller gain selection based on time domain specifications of damping factor and natural frequency are provided along with simulations indicating the selectivity of process response. The utility of proposed controller structure is shown by simulating the depth control of a nonlinear autonomous underwater vehicle system by the proposed controller structure.
\end{abstract}
\section{Introduction}
The problem of controlling integrating plus dead time (IPDT) processes has received significant attention of control engineers during past few years. Integrating processes with dead time are not only interesting from a controller design point of view but also are mathematically simple, making it easy to work analytically with these systems. Numerous PID controller tuning methods for integrating processes have been proposed in the recent past \cite{mercader2017pi,huba2018extending,kadam2017disturbance,huba2017comparing,huba2018limits,huba2018pidmncontrol,huba2018introduction,huba2013performance,huba2012setpoint,Kadam2013}.
However, this is not a new interest in IPDT systems and there have been many contributions in the past couple of decades \cite{wang2,luyben,rivera,pomerleau,camacho,chidambaram,zhong,sree,chidambaram,Kaya2003111}.

The approach used for the design of proposed control scheme for the  control of integrator plus dead-time class of systems is to design a regulator for the IPDT process first ignoring the presence of dead time and then to achieve setpoint tracking by means of a separate tracking controller. 
\begin{figure}[!htbp]
	\begin{center}
		\includegraphics[width=0.75\linewidth]{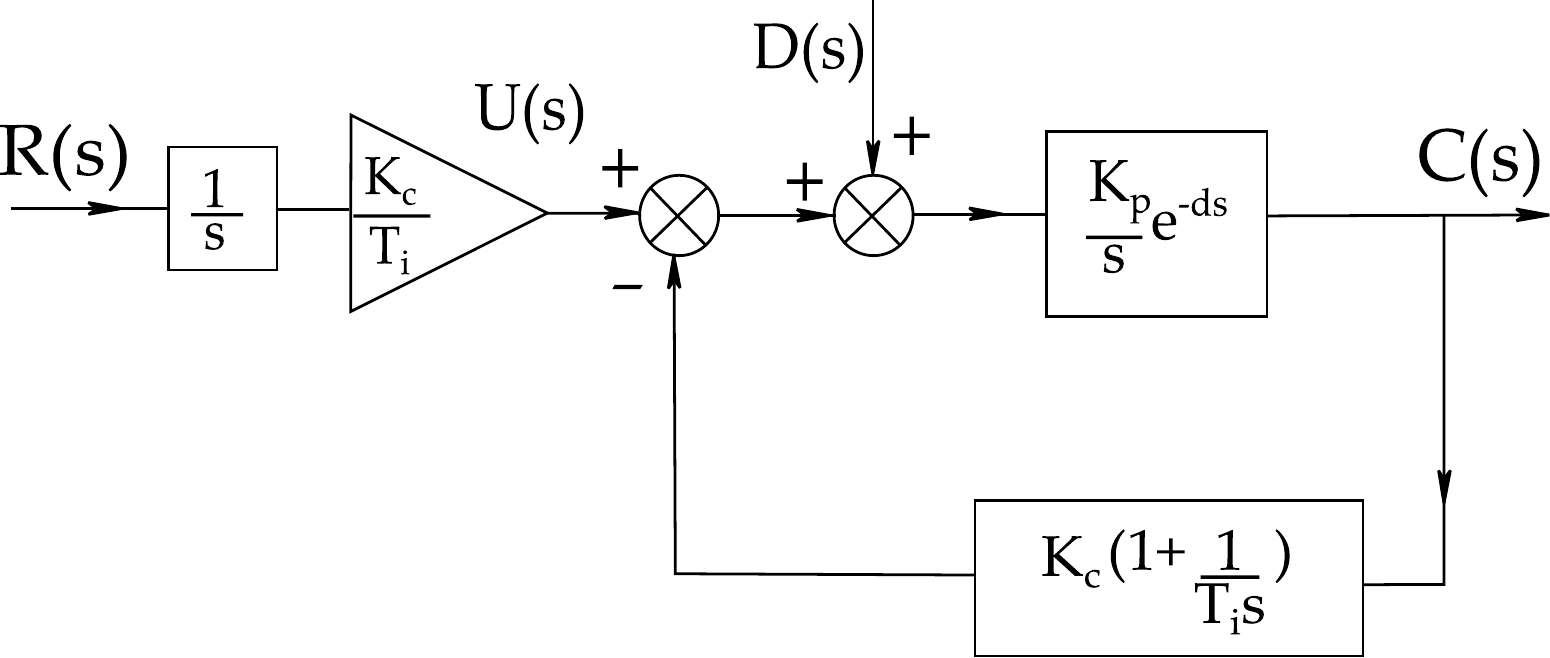}
		\caption{The I-PI controller structure}
		\label{fig:IPI_structure}
	\end{center}
\end{figure}
Fig. (\ref{fig:IPI_structure}) shows the proposed a feedback PI regulator in closed loop with the IPDT process and a feedforward I controller for setpoint tracking. The feedback compensators are known for their ability of rejecting disturbances. The technique presented here will address primarily the issue of disturbance rejection for IPDT processes. In an attempt to control the IPDT systems, for both set-point tracking and regulation, a combination of feed-forward tracking controller with a PI feedback regulator is proposed in this article. Also, the tuning of this I + PI controller scheme is suggested later in Section \ref{sec:IPI}.
\section{Derivation of PI controller settings}\label{sec:IPI}
Assume an integrating process with dead-time with transfer function given by,
\begin{equation}\label{eq:IPDT}
G(s)=\frac{K_{\rm p}}{s}e^{-ds}
\end{equation}
In (\ref{eq:IPDT}), $K_{\rm p}$ is the process gain or the integrator's gain and $d$ is the time delay element associated with it.  The proposed I + PI controller scheme is illustrated in Fig. (\ref{fig:IPI_structure}).

\begin{figure}[!htbp]
	\begin{center}
		\includegraphics[width=0.46\linewidth]{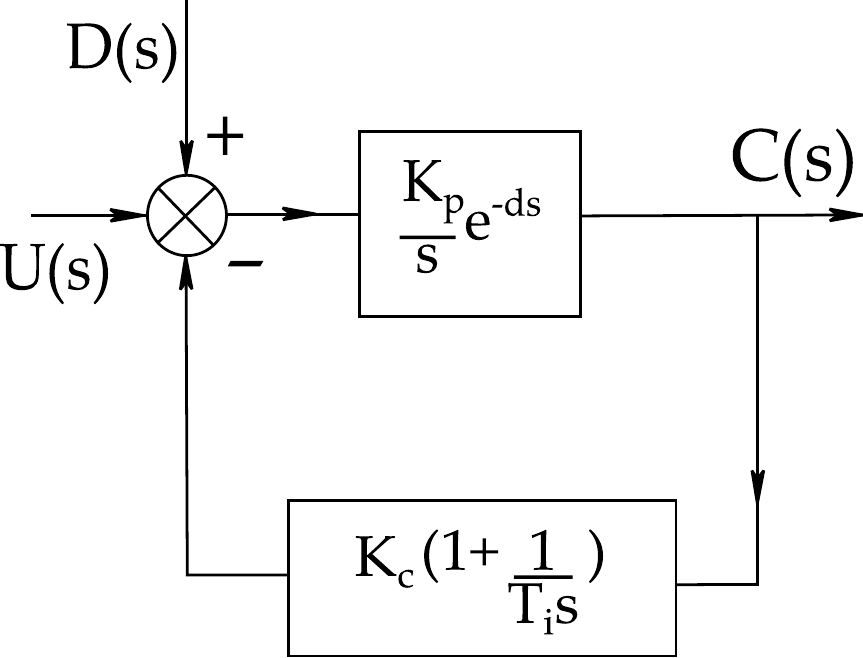}
		\caption{The I-PI controller structure}
		\label{fig:IPI_structure_d}
	\end{center}
\end{figure}

The I controller is a feed forward controller containing only the I mode of the PI feedback compensator. The transfer functions for feedforward I $(G_{\rm ff}(s))$ and feedback PI $(G_{\rm fb}(s))$ are given by,
\begin{equation}
G_{\rm ff}(s)~=~\frac{K_{\rm c}}{T_{\rm i} s}
\end{equation}
and,
\begin{equation}
G_{\rm fb}(s)~=~K_{\rm c}(1+\frac{1}{T_{\rm i} s})
\end{equation}
The block diagram shown in Fig.(\ref{fig:IPI_structure}) has been redrawn as shown in Fig.(\ref{fig:IPI_structure_d}) for deriving the controller settings for the PI feedback compensator.

The closed loop system shown in Fig.(\ref{fig:IPI_structure_d}) will have input $U(s)+D(s)$. Therefore, the closed loop transfer function for the block diagram in Fig.(\ref{fig:IPI_structure_d}) may be written as,
\begin{equation}
\frac{C(s)}{U(s)+D(s)}=\frac{\frac{K_{\rm p}}{s}e^{-ds}}{1+\frac{K_{\rm p}K_{\rm c}}{s}(1+\frac{1}{T_{\rm i} s})e^{-ds}}
\end{equation}
Assume that the dead time ($d$) is very small or negligible i.e. $d~=~0$, for simplicity.  The transfer function above will then change as,
\begin{equation}\label{eq:IPI_cltf0}
\frac{C(s)}{U(s)+D(s)}=\frac{\frac{K_{\rm p}}{s}}{1+\frac{K_{\rm p} K_{\rm c}}{s}(1+\frac{1}{T_{\rm i} s})}
\end{equation}
Simplifying,
\begin{equation}\label{eq:IPI_cltf1}
\frac{C(s)}{U(s)+D(s)}=\frac{{K_{\rm p}}{s}}{s^2+K_{\rm p} K_{\rm c} s+\frac{K_{\rm p} K_{\rm c}}{T_{\rm i}}}
\end{equation}

Re-writing above equation (\ref{eq:IPI_cltf1}),
\begin{equation}\label{eq:IPI_cltf2}
\frac{C(s)}{U(s)+D(s)}=\frac{\frac{K_{\rm p}K_{\rm c}}{T_{\rm i}}}{s^2+K_{\rm p} K_{\rm c} s+\frac{K_{\rm p} K_{\rm c}}{T_{\rm i}}} \times \frac{T_{\rm i} s}{K_{\rm c}}
\end{equation}
Now, comparing the first term in above equation with the standard closed loop transfer function for a second order system given by,
\begin{equation}\label{eq:2nd_order}
G(s)=\frac{{\omega_n}^2}{s^2 + 2 \zeta \omega_n s + {\omega_n}^2}
\end{equation}
we may have the following equations:
\begin{equation}\label{eq:zeta_Wn}
\begin{split}
2 \zeta \omega_n =K_{\rm p} K_{\rm c} \\
{\omega_n}^2 = \frac{K_{\rm p} K_{\rm c}}{T_{\rm i}}
\end{split}
\end{equation}
where, $\zeta$ is the damping factor and $\omega_n$ is the undamped frequency of oscillations.
Re-arrangement of the above equation (\ref{eq:zeta_Wn}) results into the settings for PI controller given by,
\begin{equation}\label{eq:IPI_settings}
\begin{split}
K_{\rm c} = \frac{2 \zeta \omega_n}{K_{\rm p}} \\
T_{\rm i}=\frac {2 \zeta}{\omega_n}
\end{split}
\end{equation}

From the knowledge of process gain ($K_{\rm p}$), specifications of damping factor ($\zeta$) and natural frequency of oscillation ($\omega_n$), the PI settings can be known. Alternatively, $\omega_n$ may be indirectly specified as,
\begin{equation}\label{eq:specify_Wn}
\omega_n~=~\frac{4}{(\zeta (T_s~+~d))} \times k
\end{equation}
where, $k$ is a constant multiplying factor that can select the aggressiveness of the control action.  $T_s$ is the desired settling time and this may in turn be specified by,
\begin{equation}\label{eq:specify_Ts}
T_s~=~\frac{d}{K_{\rm p}}
\end{equation}
Since, $d$ and $K_{\rm p}$ are known from the process model, $T_s$ is automatically specified. In the above equations (\ref{eq:specify_Wn}) and (\ref{eq:specify_Ts}), the dead-time term $(d)$ which was neglected previously in (\ref{eq:IPI_cltf0}), is now included while specifying $T_s$ and therefore, to specify  $\omega_n$. Thus the tuning values include the specifications of both the parameters ($K_{\rm p}$ and $d$) defining the IPDT system. The only unknown left to be chosen is the constant $k$. Higher value of $k$ will imply aggressive and faster control action. However if the value of $k$ is sufficiently high, it can result in an uncontrolled, unstable system. Also, value of damping coefficient $\zeta$ must be chosen to have an acceptable transient response.
  
From (\ref{eq:IPI_cltf2}), it can be interpreted that the system will  respond for both $U(s)$ and $D(s)$ in the same manner. So as to make the IPDT process follow the setpoint, a feedforward controller must be used such that it has a transfer function equal to the reciprocal of the second term on the right hand side of equation (\ref{eq:IPI_cltf2}). The feedforward controller then is selected to have a transfer function of the following form:
\begin{equation}\label{eq:IPI_ff}
\frac{U(s)}{R(s)}=\frac{K_{\rm c}}{T_{\rm i}} \frac{1}{s}
\end{equation}
It is obvious from above equation that the feedforward controller is an ideal form I mode only controller along with the proportional gain $K_{\rm c}$ as the multiplying factor. It must also be noted that the settings of both feedforward and the feedback controller are the same.

\begin{figure}[!htbp]
	\begin{center}
		\includegraphics[width=\linewidth]{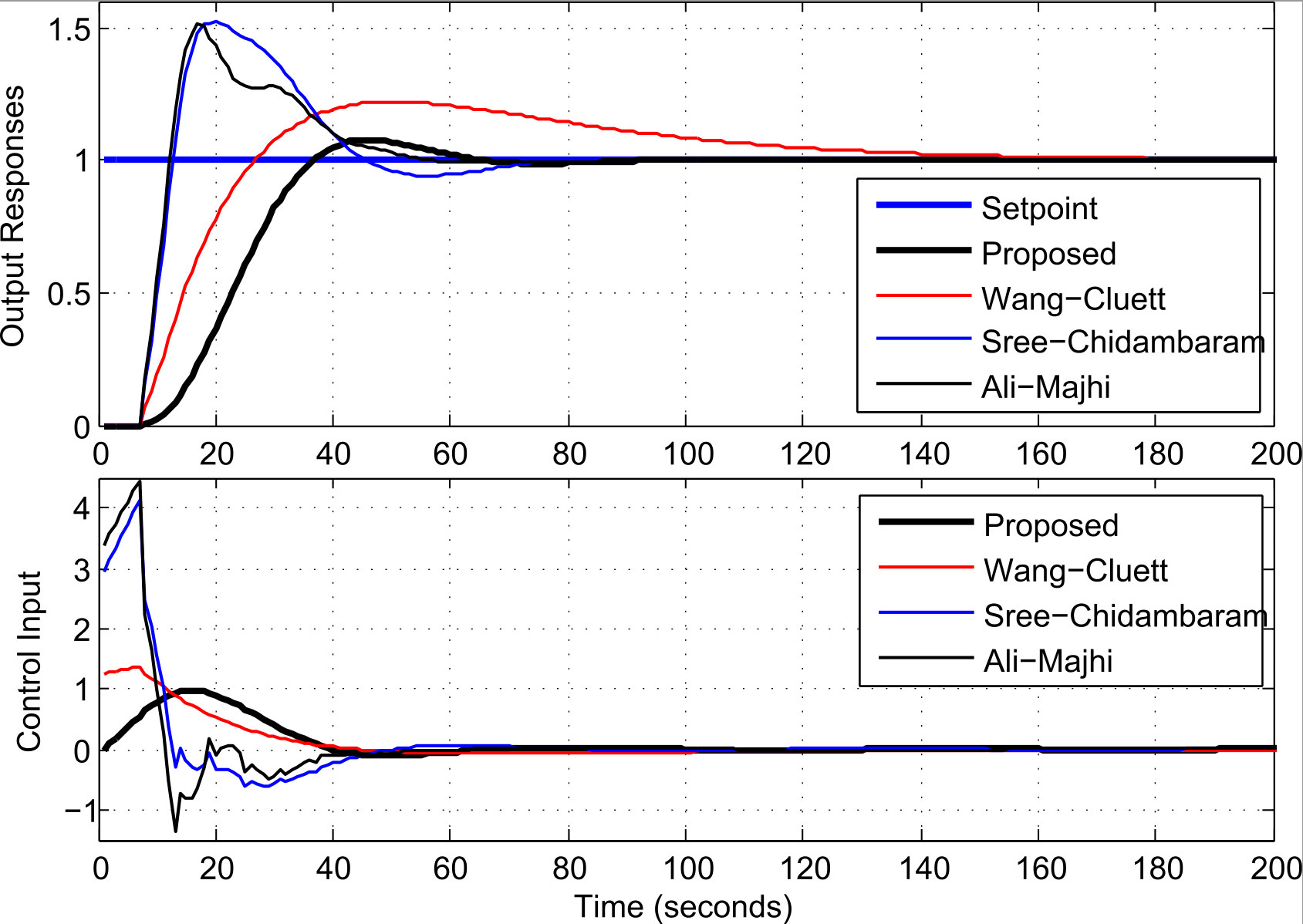}
		\caption{Setpoint tracking response obtained by the Feedforward I controller + Feedback PI compensator}
		\label{fig:IPI_tracking}
	\end{center}
\end{figure}

Consider the IPDT process given in \cite{wang2,sree,ali}  represented by the transfer function,
\begin{equation}\label{eq:IPDTsim}
G_I(s)=\frac{0.0506}{s}e^{-ds}
\end{equation}
i.e. $K_{\rm p}~=~0.0506$ and $d~=~6$.
The the plot showing setpoint tracking responses for control scheme suggested in Fig.({\ref{fig:IPI_structure}}) is shown in Fig. (\ref{fig:IPI_tracking}).

\begin{figure}
\begin{center}
\includegraphics[width=\linewidth]{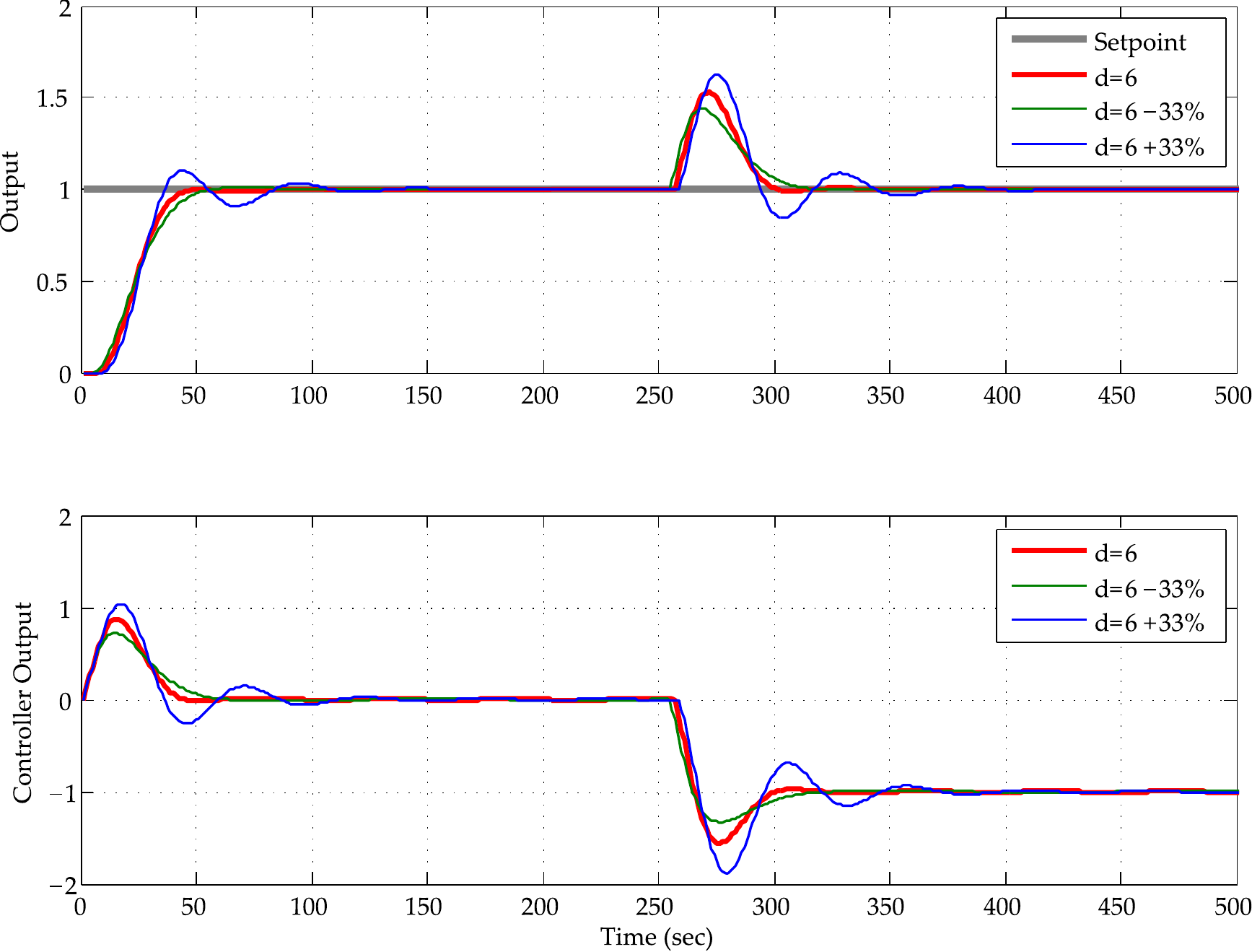}
\label{fig:IPI_output}
\caption{Effects of variation in dead-time for specified values of $\zeta$ and $\omega_n$}
\end{center}
\end{figure}
The plot also shows responses for methods presented in \cite{wang2,sree,ali}. These methods are selected for comparison with the proposed method because these methods use the standard form PID controller tuned by their respective tuning formulas for IPDT systems. The proposed method, however, does not employ the D mode of the commonly used PID, as D mode cannot be successfully used for regulatory control. Besides this, if the D mode is aggressive enough, as in the case of tuning methods suggested in  \cite{sree,ali}, then the control action is likely to be jerky. Such control action may prove detrimental to the performance of the final control element, thereby reducing its operating life. Table \ref{tab:transient_specs_IPI} shows comparison of time domain specifications for the proposed method and the methods in  \cite{wang2,sree,ali}.
 
\begin{figure}[!htbp]
	\begin{center}
		\includegraphics[width=\linewidth]{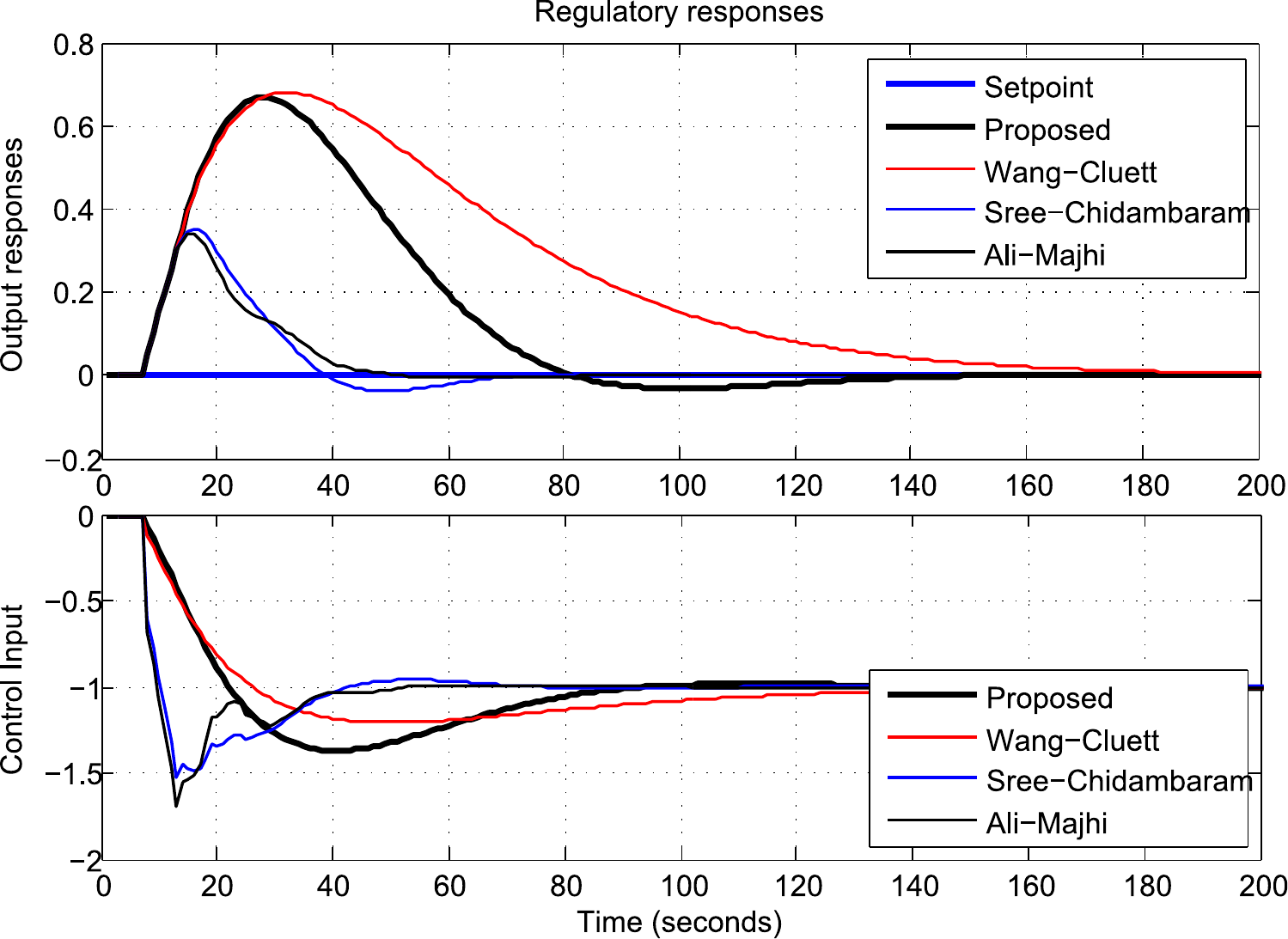}
		\caption{Comparison of regulatory responses}
		\label{fig:IPI_stepSR}
	\end{center}
\end{figure}

 \begin{table}[!htbp]
 	\caption{\textbf{Comparison of Time Domain specifications}}
 	\centering
 	\begin{scriptsize}
 	\begin{tabular}{ c  c  c  c }
	
	\hline
	\hline
	\textbf{Parameters} &\textbf{Rise Time} &\textbf{Settling Time} &\textbf{Overshoot }\\
	&(sec) &(sec) &(\%) \\
	\hline
	\hline
	\textbf{Proposed I + PI} &\textbf{18.91} &\textbf{60.10} &\textbf{7.68}\\
	\hline
	Wang - Cluett \cite{wang2} &14.63 &146.45 &22.03\\
	\hline
	Sree - Chidambaram \cite{sree} &4.49 &70.7 &52.06\\
	\hline
	Ali - Majhi \cite{ali} &4.1 &51.6 &51.51\\
	\hline
	\hline
\end{tabular}

 	\end{scriptsize}
 	\label{tab:transient_specs_IPI}
 \end{table}
  Fig. \ref{fig:IPI_stepSR} shows simulation comparisons for regulatory control.
  
\begin{figure}[!htbp]
	\begin{center}
		\includegraphics[width=\linewidth]{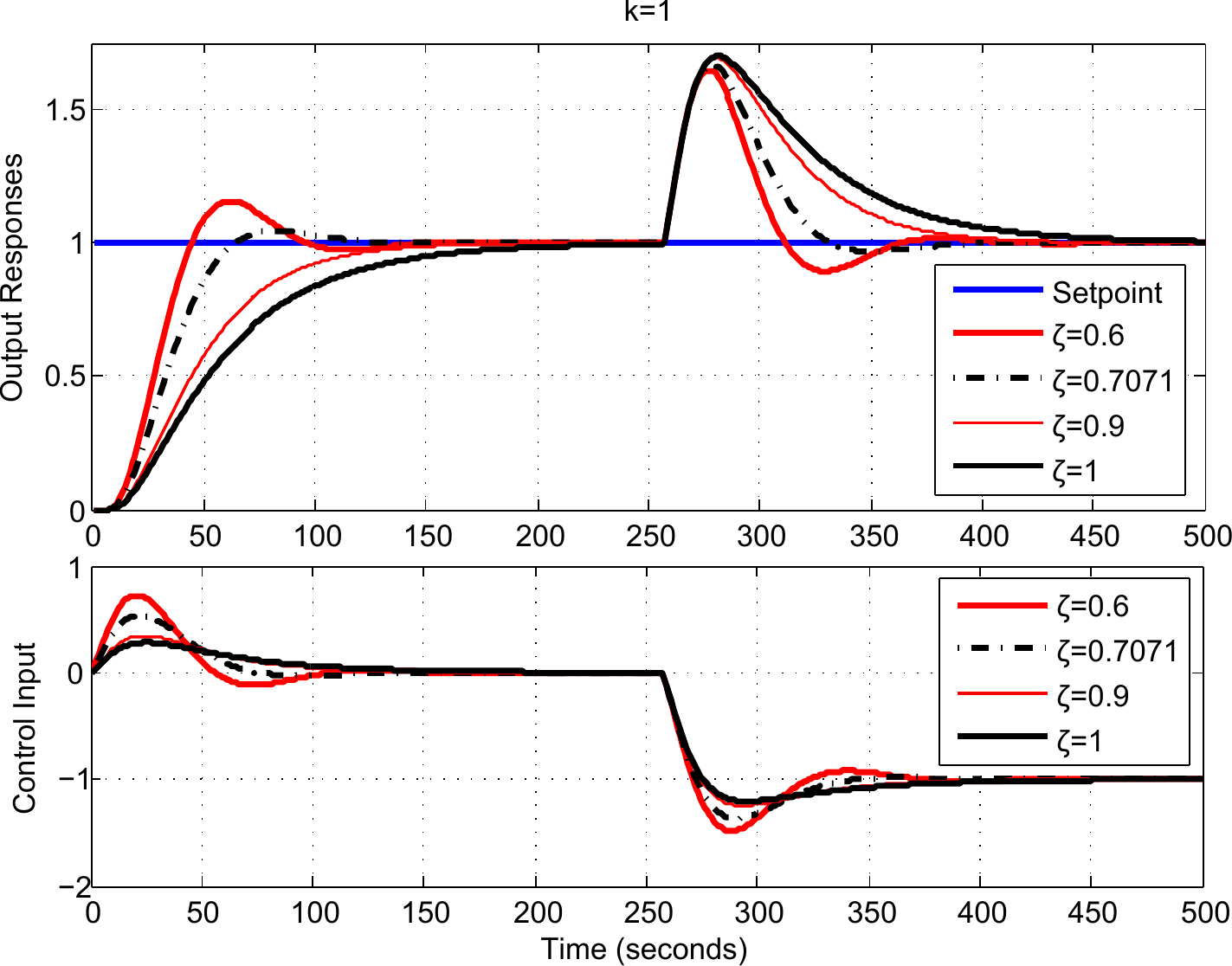}
		\caption{Responses for various values of $\zeta$ when $k$ is fixed }
		\label{fig:Wn_fixed}
	\end{center}
\end{figure}

\begin{figure}[!htbp]
	\begin{center}
		\includegraphics[width=\linewidth]{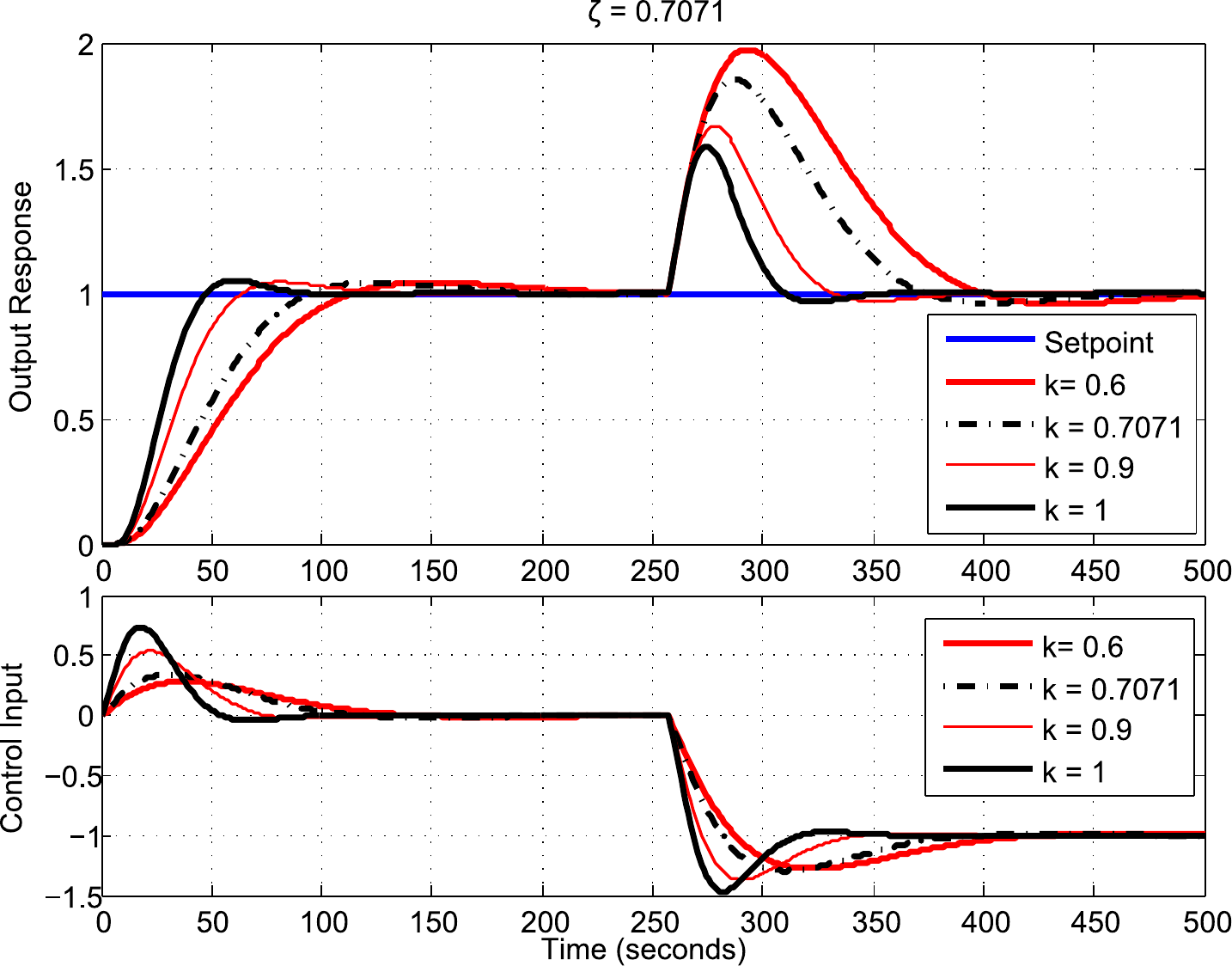}
		\caption{Responses for various values of $k$ when $\zeta$ is fixed}
		\label{fig:zeta_fixed}
	\end{center}
\end{figure}

The effects of change in the value of the specifications $\zeta$ and $\omega_n$ are demonstrated in the Figures (\ref{fig:Wn_fixed}) and (\ref{fig:zeta_fixed}) respectively. Robustness of the proposed scheme to variations in modeled dead-time values can be seen from the plots in Fig. \ref{fig:IPI_outputs}.

\section{Application to  depth control of an autonomous underwater vehicle (NPS AUV II)} \label{sec:NPSAUV}
The method for controlling IPDT systems proposed in the earlier section can be applied for the depth control of autonomous underwater vehicles. Here, the depth control of an AUV (NPS AUV II) \cite{Healey1993,fossengnc} has been discussed. The dynamics of the AUV presented in \cite{Healey1993,fossengnc} is very non-linear and parameter varying, complicated with actuator constraints. The same non-linear, parameter varying model is used for simulations of the control scheme proposed here. 

\begin{figure}[!htbp]
	\begin{center}
		\includegraphics[width=\linewidth]{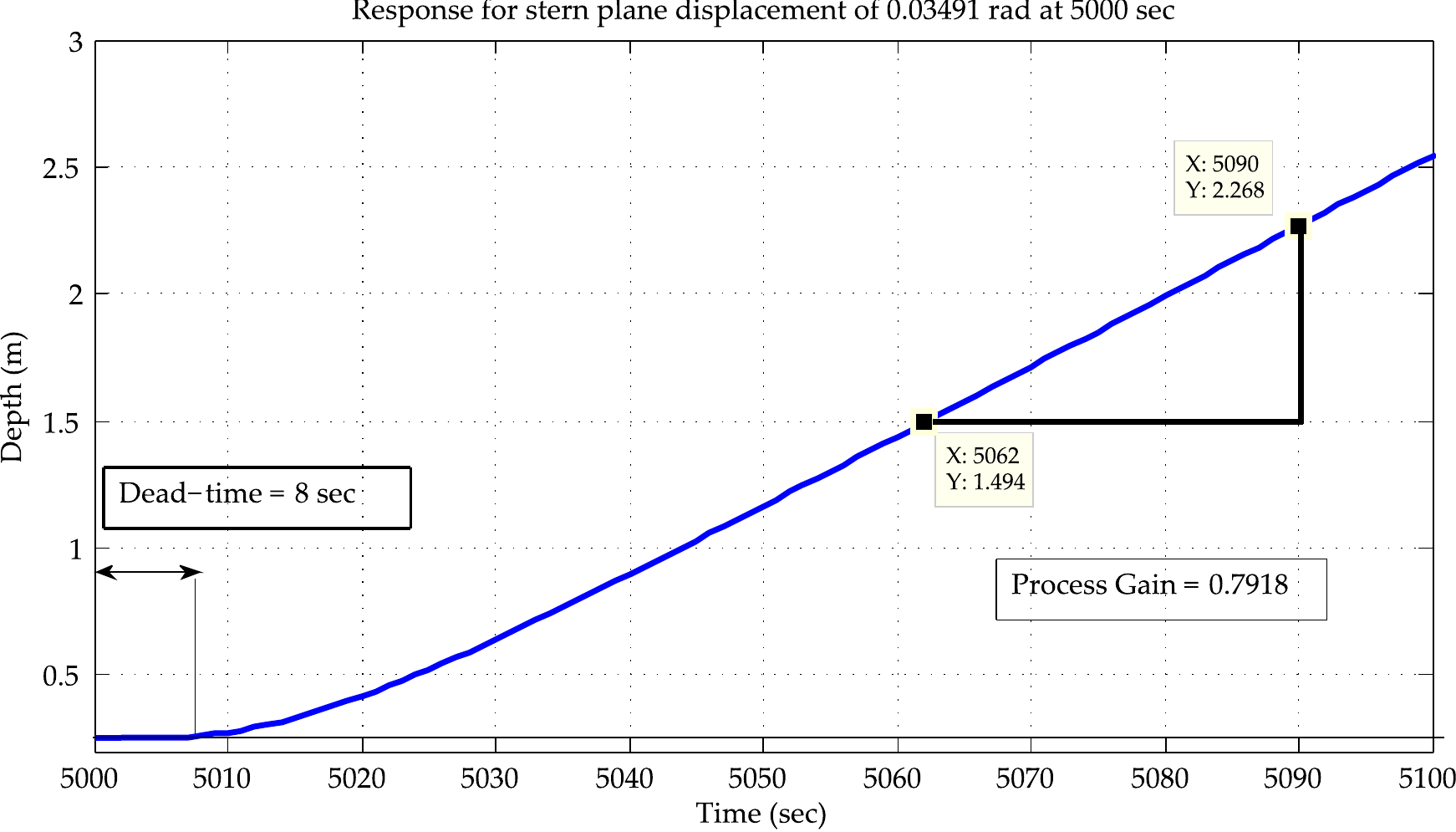}
		\caption{Approximating depth dynamics as an integrating process with dead-time}
		\label{fig:npsdepth}
	\end{center}
\end{figure}

The depth dynamics is initially modeled as an integrating process with dead time. Assuming that the AUV moves forward with a constant surge velocity $(u=0.8m/s)$, a step change of $0.03491$ rad ($2~ \deg$) is simulated at the stern planes of the AUV. Then the resulting change in the depth of the AUV is plotted. The dead-time and the process gain are then determined from the plot as shown in the Fig. (\ref{fig:npsdepth}).

\begin{figure}[!htbp]
	\begin{center}
		\includegraphics[width=\linewidth]{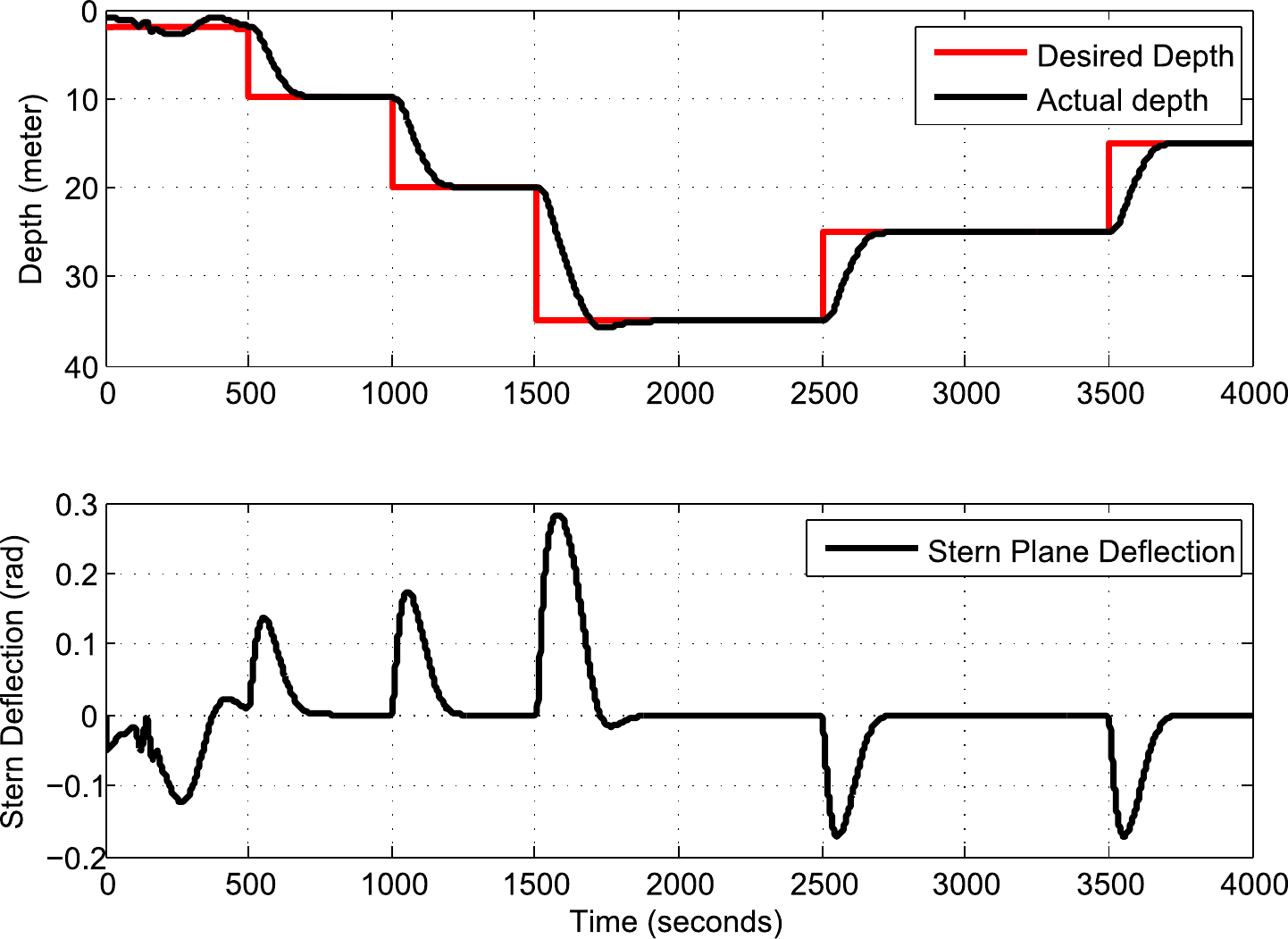}
		\caption{Depth responses and corresponding stern plane displacement}
		\label{fig:depth_PI}
	\end{center}
\end{figure}

\begin{figure}[!htbp]
	\begin{center}
		\includegraphics[width=\linewidth]{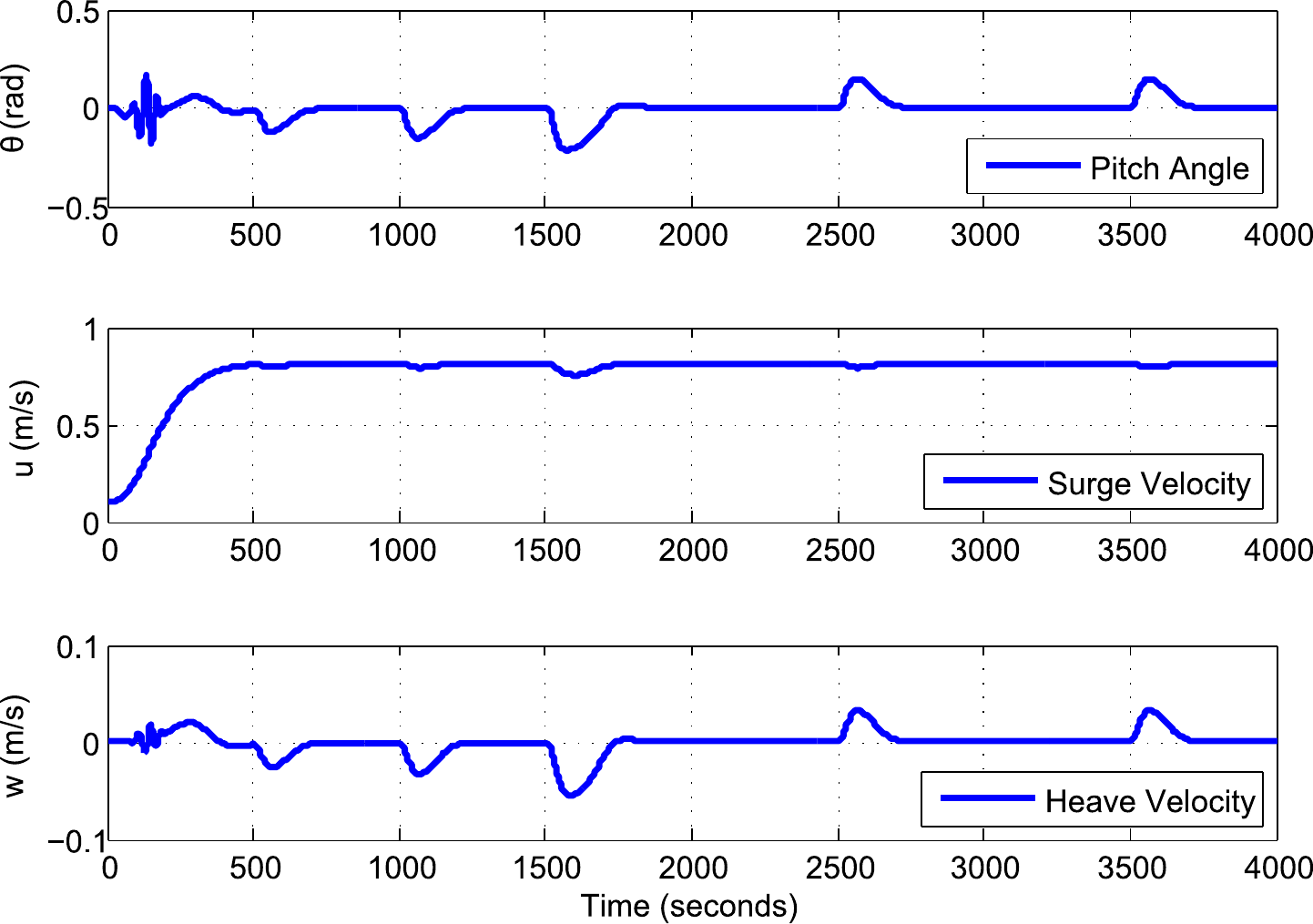}
		\caption{Other parameters comprising the depth dynamics of the AUV}
		\label{fig:thetaqw}
	\end{center}
\end{figure}

From the knowledge of process gain $K_{\rm p}=0.7918$ and the specifications of $\omega_n=0.03$ and $\zeta=0.7$ into equation (\ref{eq:IPI_settings}) result into the PI controller settings. These settings are used in the control configuration suggested in section \ref{sec:IPI}. The results for the depth control of NPS AUV II are shown in Fig.(\ref{fig:depth_PI}) along with the actuator profiles. Also the plots showing change in other parameters governing the AUV depth dynamics are plotted in Fig. (\ref{fig:thetaqw}).

\balance
\section{Conclusion}
This work presented a simple controller structure with a feedback PI and a feedforward PI controllers for the control of an integrating process with dead time. The control action of the suggested method is smooth due to the absence of derivative mode and this is important in several control applications. Simulations for depth control of a nonlinear autonomous underwater vehicle underscores the utility of this simple control scheme that can be tuned easily to control requirements.

\end{document}